\documentclass[12pt,aps]{revtex4}
\pdfoutput=1

\newcommand{\beq}{\begin{equation}}
\newcommand{\eeq}{\end{equation}}

\usepackage{graphicx}
\begin{document}

\title{Dispersion interactions and reactive collisions of ultracold polar molecules} 

\author{Svetlana Kotochigova\footnote{E-mail:skotoch@temple.edu}}

\vspace*{0.5cm}

\affiliation{Department of Physics, Temple University, Philadelphia, PA 19122-6082, USA}

\begin{abstract}
Progress in ultracold experiments with polar molecules requires 
a clear understanding of their interactions  and reactivity
at ultra-low collisional energies.  Two important 
theoretical steps in this process are the characterization of interaction 
potentials between molecules and  the modeling of 
reactive scattering mechanism.  Here, we report on the {\it ab~initio}
calculation of isotropic and anisotropic van der Waals interaction potentials 
for polar KRb and RbCs  colliding with each other or with ultracold atoms. 
Based on these potentials and two short-range scattering parameters we 
then develop a single-channel scattering model with flexible boundary conditions. 
Our calculations show that at low temperatures (and in absence of an external electric
field) the reaction rates between molecules or molecules with atoms
have a resonant character as a function of the short-range parameters.
We also find that both the isotropic and anisotropic van der Waals coefficients have
significant contributions from dipole coupling to excited electronic
states. Their values can differ dramatically from those solely obtained
from the permanent dipole moment. A comparison with recently obtained 
reaction rates of fermionic $^{40}$K$^{87}$Rb shows that the experimental data can not
be explained by a model where the short-range scattering parameters are
independent of the relative orbital angular momentum or partial wave.
\end{abstract}

\pacs{34.20.Gj,34.50.Lf,34.50.Cx,71.20.Dg}

\keywords{ultracold polar molecules, van der Waals interaction, reactive collisions, 
inelastic rates, KRb, RbCs} 

\maketitle

\section{Introduction}

Recent advances in creating ultracold polar molecules in the lowest rovibrational
ground state \cite{DeMille,Kang-Kuen} have created a new scientific
playground for studying quantum phenomena that govern collisions and interactions
between molecules at low temperatures. Some key theoretical insights into the dipolar
character of  these interactions have already been developed over the last few years
\cite{Santos,Goral,Ticknor,Krems,Avdeenkov,Ronen,Micheli}.  It was shown that
dipole-dipole forces can be quite strong and may give rise to complex many-body
physics in cold polar gases.  Molecules with dipole moments, which interact via very
long-range dipolar forces when oriented in an external electric field, can form
strongly correlated condensed matter-like systems, realize effective spin models, and
create field-linked states \cite{Micheli,Lewenstein,Avdeenkov1,Avdeenkov2}.

Currently, there is also significant interest in investigations of scattering
properties of ultracold molecules. It is expected that such properties will be very
different when molecules are held in optical dipole traps or when held in individual
sites of an optical lattice. Holding molecules in individual sites of an optical
lattice prevents them from approaching each other  and allow
dipole-dipole forces to play a dominant role in interactions between polar molecules
\cite{Rom,Thalhammer}.  Intriguingly, in lower dimensional systems, where tight
confinement along one or more spatial directions is applied, dipole forces can be
engineered to be non-destructive or repulsive and collective effects are predicted
\cite{Buchler1,Buchler2,Gorshkov,Pupillo}. For example, Ref.~\cite{Gorshkov} has
suggested a technique that decreases inelastic collisional rates and enhances elastic
collisional rates using a combination of external electric and microwave fields in a
two-dimensional lattice.

In a three-dimensional trapping environment molecules can scatter freely.  Generally,
polar molecules are expected to be very fragile to the destructive nature of these 
collisions. There is experimental evidence of the significant role that such
collisions play in defining molecular lifetimes \cite{Hudson,Jin,Jin1,YeFaraday,Science2010}. 
The physical origin of this loss are rovibrational relaxations and reactive collisions, 
both of which occur at short-range, when the molecules are close together. An understanding and quantitative
description of these collisions might help define conditions under which an ultracold
molecular system is long lived. Furthermore, learning about the short-range region will
help to unravel the role of reactivity in ultracold collisions.

The effect of reactivity  on the molecular lifetime at ultracold temperatures, with a few exceptions
\cite{Science2010,MQDT2009}, remains largely unexplored. References \cite{Weck,Hutson,Krems1,Krems2}
are devoted to theoretical developments towards a full quantum dynamics calculation of reactive collisions 
with cold molecules. The quantum mechanical description of reactions is
challenging due to the complex nature of the differential equations and boundary
conditions as well as difficulties with generating potential surfaces of sufficient
quality for three and four-atomic systems. The most successful theories are
associated with light atom-diatom systems \cite{Aldegunde,Alvarino,Tscherbul} that
are of astrophysical interest, can be Stark decelerated, or cooled with a buffer gas.
Recently, extensive quantum scattering calculations in
Refs.~\cite{Weck,Balakrishan,Bodo} have shown significant influence of so-called
 `virtual states'  in the entrance channel of the collision complex.

The interaction of molecules without an external  electric field is dominated by 
the dispersive van der Waals forces rather than dipole-dipole interactions.  At 
low temperatures quantum mechanical effects play a
prominent role in the molecular scattering from such potentials and are
crucial for the description of the interplay between inelastic and elastic
collisional rates. The present study explores  these dispersion interactions, which act at short- to
medium-range, and introduces a scattering model of ultracold polar alkali-metal
molecules. We use a characteristic length scale to distinguish between
the dispersion and short-range interaction regions. The short-range boundary, which defines the lower
limit of the scattering model, is described by 
$C^{\rm iso}_6/R_{\rm sr}^6 \ll 2 B_e$, where $C^{\rm iso}_6/R_{\rm sr}^6 $ is the isotropic dispersion
potential and $B_e$ is the rotational constant of the ground state molecule. The short-range separation 
$R_{\rm sr}$ must be smaller than  the isotropic van der Waals length $R_6 = (2\mu C^{\rm iso}_6/\hbar^2)^{1/4}$,
where $\mu$ is the reduced mass of the molecules.  
An accurate estimate of  these characteristic length scales for KRb and RbCs will be given below.

Our scattering model is an extension of the theory developed by Mies 
and Julienne \cite{Mies,Julienne}. 
Here we modified this scattering theory by making the short-range
boundary conditions more flexible by assuming that not all molecules
that penetrate to short-range will be lost.  This allows us to introduce
two short-range parameters linked to the rovibrational structure of exit
reaction channels.  Recently, a model based on quantum defect theory
has been introduced in Ref.~\cite{MQDT2009}, which accounts for the short-range
interactions by connecting a complex valued or optical potential to the van der Waals potential.

Since our scattering theory is solely based on 
the dispersion interaction potentials, we first  calculate 
molecule-molecule and atom-molecule van der 
Waals coefficients by integrating products of the dynamic polarizabily 
over imaginary frequencies.
Unlike for atom-atom interactions molecular interactions depend on
their rotational and vibrational state and can have contributions
from  transitions within the ground state potential as well as from
the electronically excited spectrum. For polar molecules the former
contribution is nonzero, even for levels with a well defined angular
momentum while the excited-state contribution can be significant. The excited contribution
is often  missing in the analyses of short-range molecular interactions
\cite{Demler,Gorshkov}.

The paper is organized as follows.  In section \ref{theory} we describe
the theory of dispersion interaction isotropic and anisotropic potentials
between molecules and molecules with atoms.  The numerical potentials
and van der Waals coefficients are obtained for the X $^1\Sigma^+$ ground
state of KRb and RbCs as a function of its vibrational quantum number and
presented in Section \ref{C6}.  Section \ref{scatter}  gives basics of
our scattering model and provide examples of the inelastic scattering
rates for ultracold colliding molecules. The last Section \ref{last}
summarizes essential conclusions of the presented research.

\section{Dispersion potentials for the molecular interaction} \label{theory}

We describe the dispersion interaction potential between molecules $A$
and $B$, each in a rovibrational level $|{\rm X}, v J M \rangle\equiv
|i,M\rangle$ of their X electronic ground state, by assuming that the
molecules are far apart and their wavefunctions  do not overlap.  Here,
the magnetic quantum number $M$ is the projection along a laboratory
fixed coordinate system of their total angular momentum $\vec{J}$
and $i$ describes all other quantum labels.  Their energy $E_i$ only
depends on the label $i$ and not on $M$.  These assumptions allow us
to use a (degenerate) second-order perturbation theory similar to that described
by Ref.~\cite{Stone}.  For two molecules the matrix elements of
the dispersion potential between the product states $|i_A,M_A;
i_B,M_B\rangle\equiv |i_A,M_A\rangle
|i_B,M_B\rangle $  and $|i_A,M'_A; i_B,M'_B\rangle$ with different projection quantum numbers 
but the same angular momenta $J_A$ and $J_B$ are
\begin{eqnarray} 
   U_{\rm disp}(\vec R) &=&
        -  \sum_{S_A\neq i_A,M_A\atop S_B\neq i_B,M_B}  
               \frac{  \langle i_A,M_A; i_B,M_B | V_{dd} | S_A; S_B \rangle 
        \langle S_A; S_B | V_{dd} | i_A M'_A; i_B,M'_B\rangle }
                  { E_{S_A} + E _{S_B}- (E_{i_A} + E _{i_B}) } \,,
\label{disp}                
\end{eqnarray}
where the sums $S_A$  and $S_B$ are over all electronic,  rovibrational,
continuum states of molecules $A$  and $B$, respectively, restricted to
states with energy $E_{S_A} + E _{S_B}\neq E_{i_A} + E _{i_B}$.
The operator $V_{dd}$ is the dipole-dipole interaction Hamiltonian \cite{Stone}
\begin{equation}
    V_{dd}(\vec R)  =  - \sqrt{30}\sum_{m_1m_2m} 
                \left( \begin{array}{ccc} 1 & 1 & 2 \\ m_1 & m_2 & m \end{array} \right)
                    d^A_{1m_1} d^B_{1m_2}  \frac{C_{2m}(\hat R)}{R^3}\,,
\label{dipole}                    
\end{equation}
where $d^A$ and $d^B$ are the two rank-1 spherical dipole operators
of the molecules and $\vec R$ is the separation between and
orientation of the two molecules with respect to a laboratory axis.
The $({\cdots\atop\cdots})$ is a Clebsch-Gordan coefficient and
$C_{lm}(\hat R)$ is the modified spherical harmonic function \cite{Brink}.
Inserting Eq.~(\ref{dipole}) into Eq.~(\ref{disp}) and  after a number of
transformations following Ref.~\cite{Stone} we find that the dispersion
potential is the sum of the isotropic $U_{\rm disp,iso}(\vec R)$
and anisotropic $U_{\rm disp,aniso}(\vec R)$ potentials, which can be
expressed in  terms of the molecular dynamic polarizability tensor at
imaginary frequency, $\alpha^{A/B}(i\omega)$, of molecule $A$ and $B$.

We find an isotropic $U_{\rm disp,iso}(\vec R)$ potential 
\begin{eqnarray}
  U_{\rm disp,iso}(R) & = & - \frac{C^{\rm iso}_6}{R^6} 
             \delta_{M_A,M'_A}  \delta_{M_B,M'_B}
               -  \frac{C^{\rm iso}_{6,22}}{R^6} \langle i_A, M_A; i_B M_B| T_{0,0}(2,2) |i_A, M'_A; i_B M'_B\rangle \,,
\label{isopot}
\end{eqnarray}
where the rank-$l$ tensor operator $T_{l,m_l}(k,p)$ is defined by
\begin{eqnarray}
 \langle i_A, M_A; i_B, M_B | T_{l,m_l}(k,p) |i_A, M'_A; i_B, M'_B\rangle &=& \frac{1}{N_l(k,p)}
    \sum_{qr}
     \langle l m_l | k p q r \rangle \label{tensor} \\
             &&\quad\quad\quad 
    \times      \langle J_A M_A | J_A k M'_A q \rangle \langle J_B M_B | J_B p M'_B r \rangle
             \nonumber
\end{eqnarray}
and the normalization
$
    N_l(k,p)= \langle l 0 | k p 0 0 \rangle
      \langle J_A J_A | J_A k J_A 0 \rangle \langle J_B J_B | J_B p J_B 0 \rangle \,,
$
such that the operator $\langle \dots |T_{lm_l}(k,p)|\dots\rangle$ is one
for the stretched state $|i_A, M_A\rangle=|i_A, M'_A\rangle= |X, v J_A J_A
\rangle$ and $|i_B, M_B\rangle=|i_B, M'_B\rangle=|X, v J_B J_B \rangle$.
Here $\langle j_3 m_3 | j_2 j_1 m_2 m_1 \rangle$ is a Clebsch-Gordan coefficient.
Consequently,
\begin{eqnarray*}
C^{\rm iso}_6 &=& \frac{3}{\pi}  \int_0^\infty d\omega 
      \langle i_A, J_A | \bar\alpha^A(i\omega) | i_A, J_A \rangle
                  \langle i_B, J_B | \bar \alpha^B(i\omega) | i_B, J_B \rangle\\
  C^{\rm iso}_{6,22} &=& \frac{3}{45\pi}  \int_0^\infty d\omega   
                  \langle i_A, J_A | \Delta\alpha^A(i\omega) | i_A, J_A \rangle  
                  \langle i_B, J_B | \Delta \alpha^B(i\omega) | i_B, J_B \rangle
\end{eqnarray*}
where, for each molecule $A$ and $B$, $\bar\alpha=( \alpha_{xx}
+ \alpha_{yy}+\alpha_{zz})/3$ and $\Delta\alpha=\alpha_{zz} -
(\alpha_{xx}+\alpha_{yy})/2$ are given in terms of diagonal $x$, $y$,
and $z$ components of the polarizability tensor at imaginary frequency.
For atoms and diatoms the polarizability tensor is fully determined by $\bar\alpha$
and $\Delta\alpha$.

The anisotropic van der Waals potential is defined by
\begin{eqnarray*}
U_{\rm disp,aniso}(\vec R) &=&
   \frac{1}{R^6} \sum_{m_l} (-1)^{m_l} C_{2m_l}(\hat R) 
  \left\{
         C_{6,02}^{\rm aniso} \langle i_A, M_A; i_B, M_B | T_{2,-m_l}(0,2) |i_A,M'_A; i_B,M'_B \rangle
  \right. \\
   &&  
   \phantom{\frac{1}{R^6} \sum_{m_l} (-1)^{m_l} C_{2m_l}(\hat R) }
       +  C_{6,20}^{\rm aniso} \langle i_A, M_A; i_B, M_B | T_{2,-m_l}(2,0) |i_A,M'_A; i_B,M'_B \rangle
   \\
   &&  
   \phantom{\frac{1}{R^6} \sum_{m_l} (-1)^{m_l} C_{2m_l}(\hat R) }
           \quad
      \left.
       + C_{6,22}^{\rm aniso} \langle i_A, M_A; i_B, M_B | T_{2,-m_l}(2,2) |i_A,M'_A; i_B,M'_B \rangle
  \right\}
\end{eqnarray*}
and the van der Waals coefficients are
\begin{eqnarray*}
   C_{6,02}^{\rm aniso} &=&
    -    \frac{1}{\pi} \int_0^\infty d\omega 
                 \langle i_A, J_A | \bar\alpha^A(i\omega) | i_A, J_A \rangle
                  \langle i_B, J_B | \Delta\alpha^B(i\omega) | i_B, J_B \rangle \,,   \\
   C_{6,20}^{\rm aniso} &=&
    -    \frac{1}{\pi} \int_0^\infty d\omega  
                   \langle i_A, J_A | \Delta\alpha^A(i\omega) | i_A, J_A \rangle
                  \langle i_B, J_B | \bar\alpha^B(i\omega) | i_B, J_B \rangle \,,   \\
   C_{6,22}^{\rm aniso} &=& \frac{2}{7} C_{6,22}^{\rm iso} \,.
\label{anisopot}                  
\end{eqnarray*}

In our framework the $x, y$, and $z$ component of the diagonal dynamic
polarizability  at imaginary frequency is determined as 
\beq
\langle i, M |\alpha_{nn}(i\omega)| i, M\rangle =  \frac{1}{\epsilon_0c}
   \sum_{S\neq i, M} \frac{(E_S -  E_i)}{(E_S  - E_i)^2 - (i\hbar\omega)^2}
     \times |\langle S|\vec{d}\cdot \hat{n}|i, M\rangle|^2 \,
\label{eqpolar}
\eeq
where $\hat{n}$ is the unit vector along the $n$=$x$, $y$, and $z$
direction, and $i, M$ and $S$ denote rovibrational wave functions of a single
molecule, $\langle S|\vec d|i, M \rangle$ are matrix elements of permanent
or transition electronic dipole moments.  Equation~(\ref{eqpolar})
includes a sum over the dipole transitions to the rovibrational levels
within the ground-state potential as well as to the rovibrational levels
of electronically-excited potentials.  Contributions from scattering or
continuum states of the electronic potentials are also included.  Finally,
$c$ is the speed of light and $\epsilon_0$ is the electric constant.

In summary, we have found various contributions to the long-range van der Waals potential.
They are characterized by their angular momentum dependence. In particular,
the term proportional to $C^{\rm iso}_6$ is independent of magnetic projections
and the relative orbital angular momentum between the molecules.
The term proportional to $C^{\rm iso}_{6,22}$  induces coupling between the 
magnetic projections of the two molecules without affecting the relative orbital angular momentum.
Finally, the three $C^{\rm aniso}_6$ contributions do cause mixing between magnetic projections
and relative orbital angular momentum.
Note that the coefficients $C^{\rm iso}_{6,22}$ and $C^{\rm aniso}_{6,22}$ will be of
the same order of magnitude. They both are proportional to a product of two $\Delta\alpha$.
Dispersion terms proportional to the spherical harmonic $C_{lm}(\hat R)$ for $l>2$
also exist.  We do not consider them here as they are smaller.

\section{Van der Waals coefficients of polar molecules} \label{C6}

This section describes our results for the van der Waals coefficients $C^{\rm iso}_6$ and 
$C^{\rm aniso}_6$ between two polar KRb dimers and two RbCs dimers. These  
molecules are prepared in rovibrational states of the ground X$^1\Sigma^+$ potential  
and  are of interest for on-going ultracold experiments \cite{DeMille,Kang-Kuen,Science2010,Pilch}.
We have used a non-relativistic {\it ab~initio} version of the electronic
structure multi-reference configuration interaction method
\cite{KotochigovaRbCs,KotochigovaSr2} to obtain potential energies,
permanent and transition electric dipole moments of the KRb and RbCs
molecules as a function of internuclear separation. In order to determine
experimental observables we combine our electronic structure calculations
with that of the calculation of rotational-vibrational energy levels to
find vibrationally-averaged transition dipole moments.

\begin{figure}
\includegraphics[scale=0.3,viewport=0 20 750 530,clip]{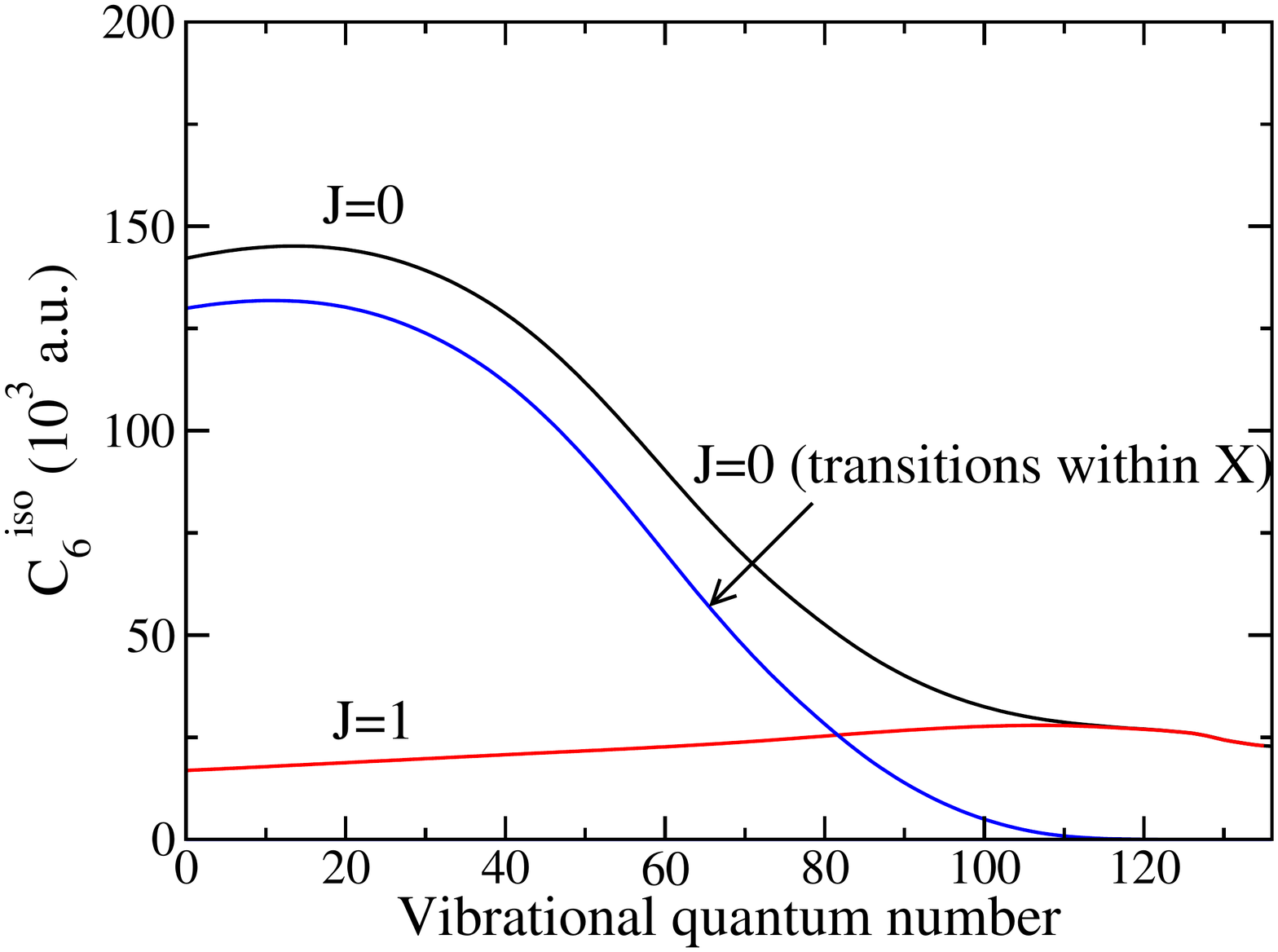}\ 
\includegraphics[scale=0.3,viewport=0 20 750 530,clip]{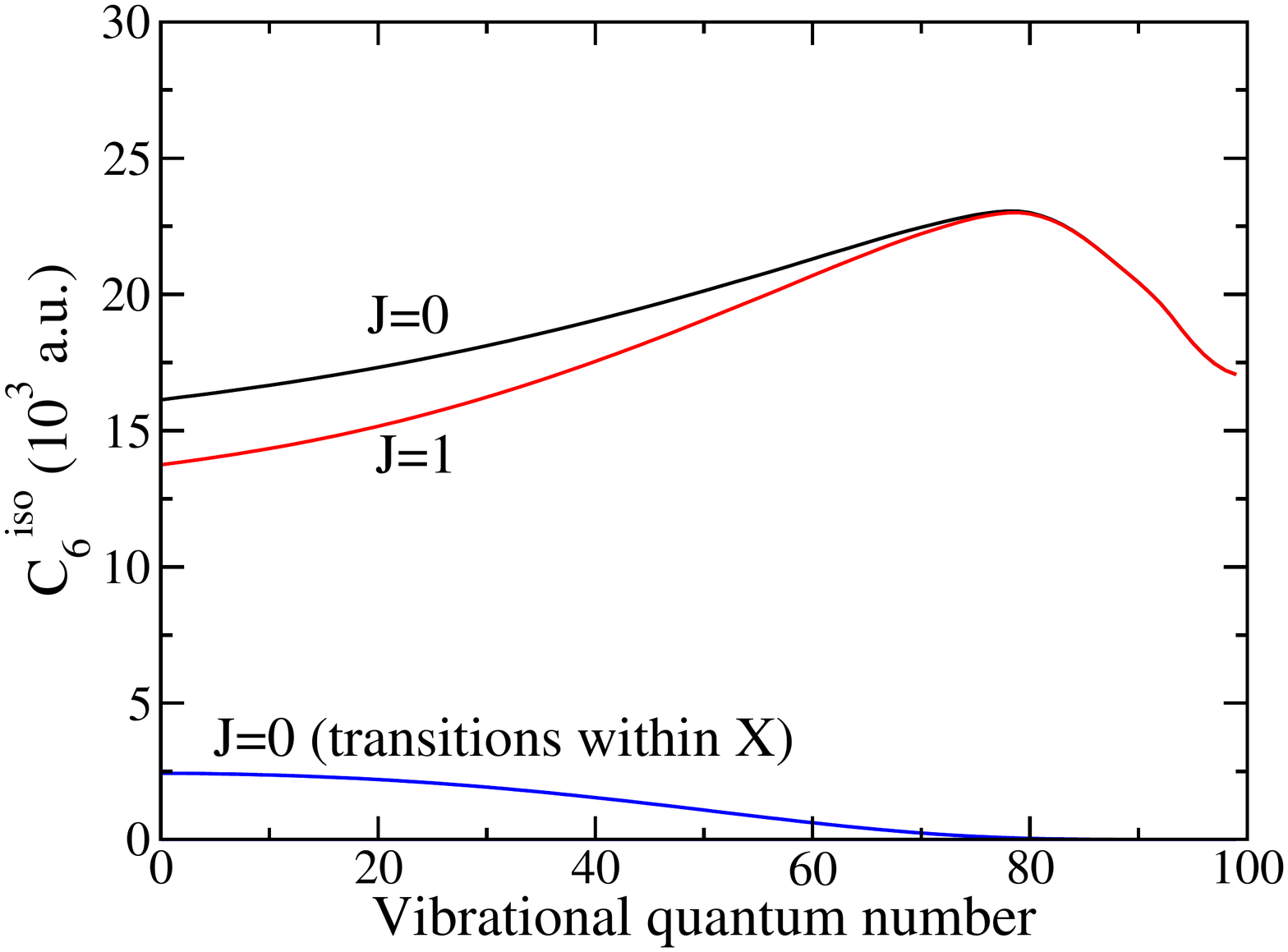}
\caption{Isotropic molecule-molecule van der Waals coefficients in atomic units for the $J$=0 
and 1 rotational levels of the X$^1\Sigma^+$ ground state of RbCs  (left panel) and KRb 
(right panel)  as a function of vibrational quantum number.
The curves labeled ``$J=0$ (transitions within X)'' correspond to an isotropic  van der Waals 
coefficient where only contributions from transitions to ro-vibrational levels of  the X$^1\Sigma^+$  potential are included.}
\label{C6rbcs_krb}
\end{figure}

The isotropic $C^{\rm iso}_6$ coefficients between two molecules both
in the $J$=0 rotational levels  of  the X $^1\Sigma^+$ ground state
potential as a function of vibrational quantum number are shown in
Fig.~\ref {C6rbcs_krb}.  Comparison of the two panels show that for
the deeply bound $J$=0 vibrational  levels the isotropic coefficient
for RbCs is almost an order of magnitude larger than for KRb. This is
because for these levels the main contribution to the total value of
$C^{\rm iso}_6$ is due to transitions within the ground state potential,
which to good approximation is equal to $d_v^4/(6B_v)$ \cite{Demler},
where $d_v$ and $B_v$ are the vibrationally averaged permanent dipole
moment and rotational constant of level $v$, respectively. The permanent
dipole moment of RbCs is twice as large as that of KRb, while $B_v$ is
almost half the size \cite{KotochigovaKRb,Kang-Kuen,KotochigovaRbCs}.
The permanent dipole moment of both molecules rapidly decreases to
zero with $v$  and the X $^1\Sigma^+$  contribution to the van der
Waals coefficient follows this trend.  Consequently, for the highly
excited vibrational levels the $C^{\rm iso}_6$ are solely determined
by transitions to electronically excited states.  In fact, for KRb the
excited state contribution plays a dominant role for all $v$.

Figure~\ref{C6rbcs_krb} also shows the $J$=1 isotropic dispersion
coefficients. These are independent of the projection quantum number $M$.
In this case the contributions from the transitions from $J$=1 to $J$=0
and 2 within the ground state potential have opposite sign and nearly
cancel each other. As a result, the value of $C^{\rm iso}_6$ is close
or equal to the value obtained by only including  the excited state
contributions.  The dispersion coefficient is weakly dependent on $v$.
The isotropic values for the $v=0$ vibrational level of the X$^1\Sigma^+$ state
of KRb and RbCs are given in Table~\ref{summary}.

The $J$=0 and 1 dispersion coefficients are the same when they are solely
determined by transitions to electronically excited states.  This is
because rotational energy splittings are negligible compared to electronic
excitation energies and the rotational energy dependence of $E_{S_A}$
and $E_{S_B}$ in Eq.~\ref{disp} can be neglected.  Alternatively,
this independence follows from a simple atomistic model for the dispersion
coefficient of weakly bound molecules.  The molecular electronic wave
functions are then to good approximation products of atomic electronic
wave functions and the molecule-molecule dispersion coefficient reduces
to a sum of atom-atom dispersion coefficients, which do not depend on
the rotational state of the molecule.  In fact, for RbCs+RbCs we find
that $C^{\rm iso}_6\to   2 C_6({\rm RbCs}) + C_6({\rm Rb}_2) + C_6({\rm
Cs}_2) = 22868$ a.u.  for large $v$ using the well-characterized Rb+Rb, 
Rb+Cs, and Cs+Cs dispersion coefficients \cite{Derevianko_C6}. This
value is in good agreement with our results.
A similar agreement is found for $C^{\rm iso}_6$ of the interacting KRb 
molecules based on values of $C_6$ from \cite{Derevianko_C6}.

\begin{figure}
\includegraphics[scale=0.28]{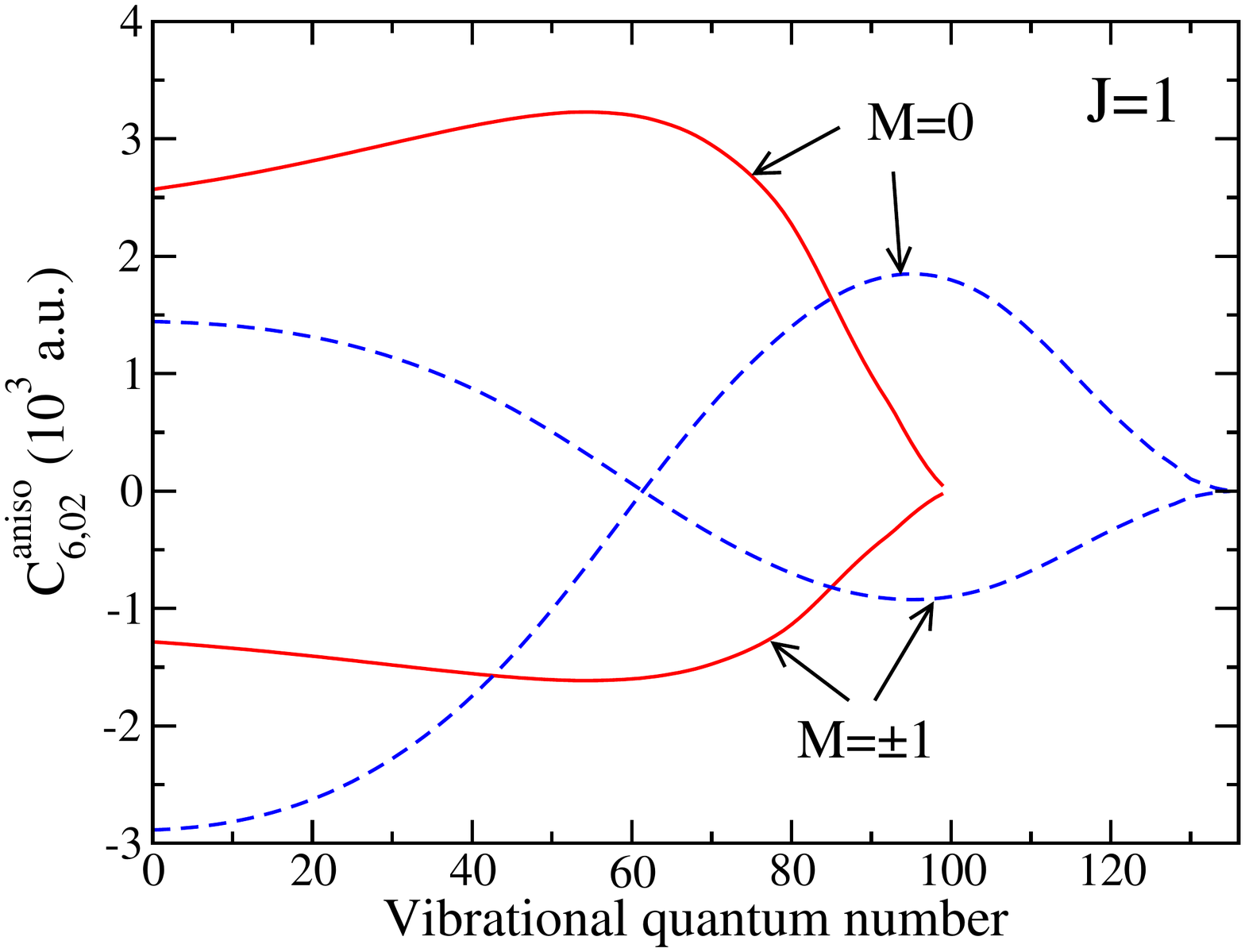}
\includegraphics[scale=0.28]{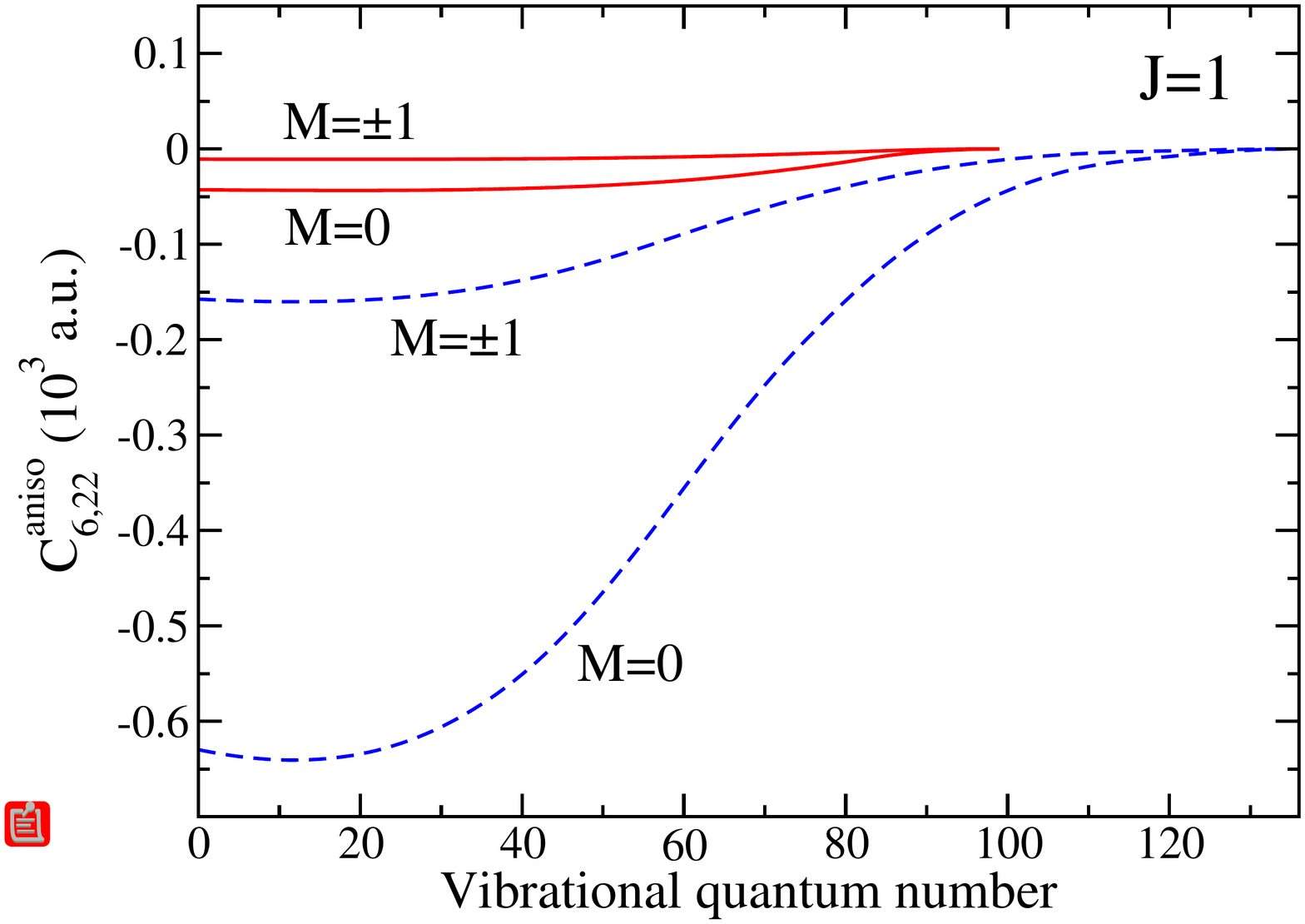}
\caption{Anisotropic van der Waals coefficients $C_{6,02}^{\rm aniso}$ (left panel)
and $C_{6,22}^{\rm aniso}$ (right panel) in atomic units as a function of vibrational quantum number for two
molecules in a $J=1$ rotational levels of the X$^1\Sigma^+$ ground state
of KRb (solid lines) and RbCs (dashed lines).  The coefficient is shown
for both $M=0$ and 1. The coefficient for $M=0$ follows from the one
for the stretched state $M=1$ by evaluation of the tensor $T_{l m_l}$
in Eq.~\ref{tensor}.}
\label{aniso}
\end{figure}

The collisions between two $J$=1 molecules are anisotropic. Once molecules
are prepared in the specific state characterized by $M$, the interaction
energy can depend on the relative orientation of the two molecules.
In fact, the anisotropic interaction  can
change the projection quantum numbers.  Figure~\ref{aniso} (left panel) shows our
results for the anisotropic $C^{\rm aniso}_{6,02}$ coefficient as a
function of vibrational quantum number for $J=1$ rotational sublevels,
$M=0$ and $M=\pm 1$, of the X$^1\Sigma^+$ ground state of KRb and RbCs.
The anisotropy for RbCs changes sign as function of $v$ due to competing
contributions from the ground and excited states. In both cases
$C_{6,02}^{\rm aniso}$ goes to zero for highly excited vibrational levels.
For these collision we have $C^{\rm aniso}_{6,20}=C^{\rm aniso}_{6,02}$.

Our predicted isotropic van der Waals $C_6^{\rm iso}$ coefficients for
the interaction between KRb molecules have been used in theoretical
models \cite{Science2010,MQDT2009} to successfully describe the loss
rate constant $\cal K$ observed in a JILA experiment as a function of
temperature \cite{Science2010}.  The anisotropic $C_6^{\rm aniso}$
coefficients can be used for describing collisions between molecules in
non-zero $J$ rotational states.  Our estimate shows that an experimentally-accesible
external electric field of 2 kV/cm  will induce an $\approx$ 50 MHz
splitting between $M=0$ and $M=\pm$ 1 components of $J=1$ rotational
state of KRb. The anisotropic interaction terms, for example, with
coefficient $C_{6,02}^{\rm aniso}\approx2500$ a.u., will then cause
a reorientation of the magnetic sublevels for separations less than 80
$a_0$ and will contribute to loss of molecules from the trap.

We also determined the coefficients $C^{\rm iso}_{6,22}$ and $C^{\rm
aniso}_{6,22}$. For rotationless $J=0$ molecules both coefficient are zero.
Right panel of Fig.~\ref{aniso} shows $C^{\rm aniso}_{6,22}$ for $J=1$ as function of vibrational 
quantum number.  These values are an order of magnitude smaller than the contributions discussed so far.
The anisotropic coefficients for both molecules in the $v=0$  X$^1\Sigma^+$ state are given in Table~\ref{summary}.

It was observed in Refs.~\cite{Hudson,Jin,Jin1,YeFaraday,Science2010} that the lifetime of both
Feshbach and deeply bound polar molecules decreases when they are in the presence of
ultracold atoms.  This suggests that collisions between atoms and
molecules play a role in limiting the molecule lifetime.  
These losses are due to relaxation of rovibrational or hyperfine
degrees of freedom that have enough kinetic energy to remove both 
molecule and atom from an external optical or magnetic trap. Another possible loss channel 
is an ultracold chemical reaction. To understand these losses,
we determine the isotropic and anisotropic van der Waals coefficients for the
interaction between a molecule in state $|i_{\rm Mol}, M_A\rangle $ and atom
in state $|i_{\rm At},M_B\rangle$. Again following Ref.~\cite{Stone}, the isotropic coefficient is
\begin{eqnarray}
  C^{\rm AtMol,iso}_6& = &  \frac{3}{\pi}   \int_0^\infty d\omega   
      \langle i_{\rm Mol}, J_A| \bar\alpha^{\rm Mol}(i\omega) | i_{\rm Mol}, J_A \rangle
      \langle i_{\rm At}, J_B| \bar\alpha^{\rm At}(i\omega) | i_{\rm At}, J_B \rangle
\,,
\label{C6atmoliso}
\end{eqnarray}
where $\bar \alpha^{\rm Mol}$ and $\bar \alpha^{\rm At}$ are
the mean molecular and atomic polarizabilities, respectively.
The anisotropic coefficient is
\begin{eqnarray}
  C^{\rm AtMol,aniso}_{6,20}& = &  \frac{1}{\pi}   \int_0^\infty d\omega   
      \langle i_{\rm Mol}| \Delta\alpha^{\rm Mol}(i\omega) | i_{\rm Mol} \rangle
      \langle i_{\rm At}| \bar\alpha^{\rm At}(i\omega) | i_{\rm At} \rangle
\,,
\label{C6atmolaniso}
\end{eqnarray}
The values for $C^{\rm AtMol,iso}_{6,22}$, $C^{\rm AtMol,aniso}_{6,02}$, and $C^{\rm AtMol,aniso}_{6,22}$
are zero as $\Delta\alpha$ is zero for an atom.
 
\begin{figure}
\includegraphics[scale=0.3,viewport=0 20 750 530,clip]{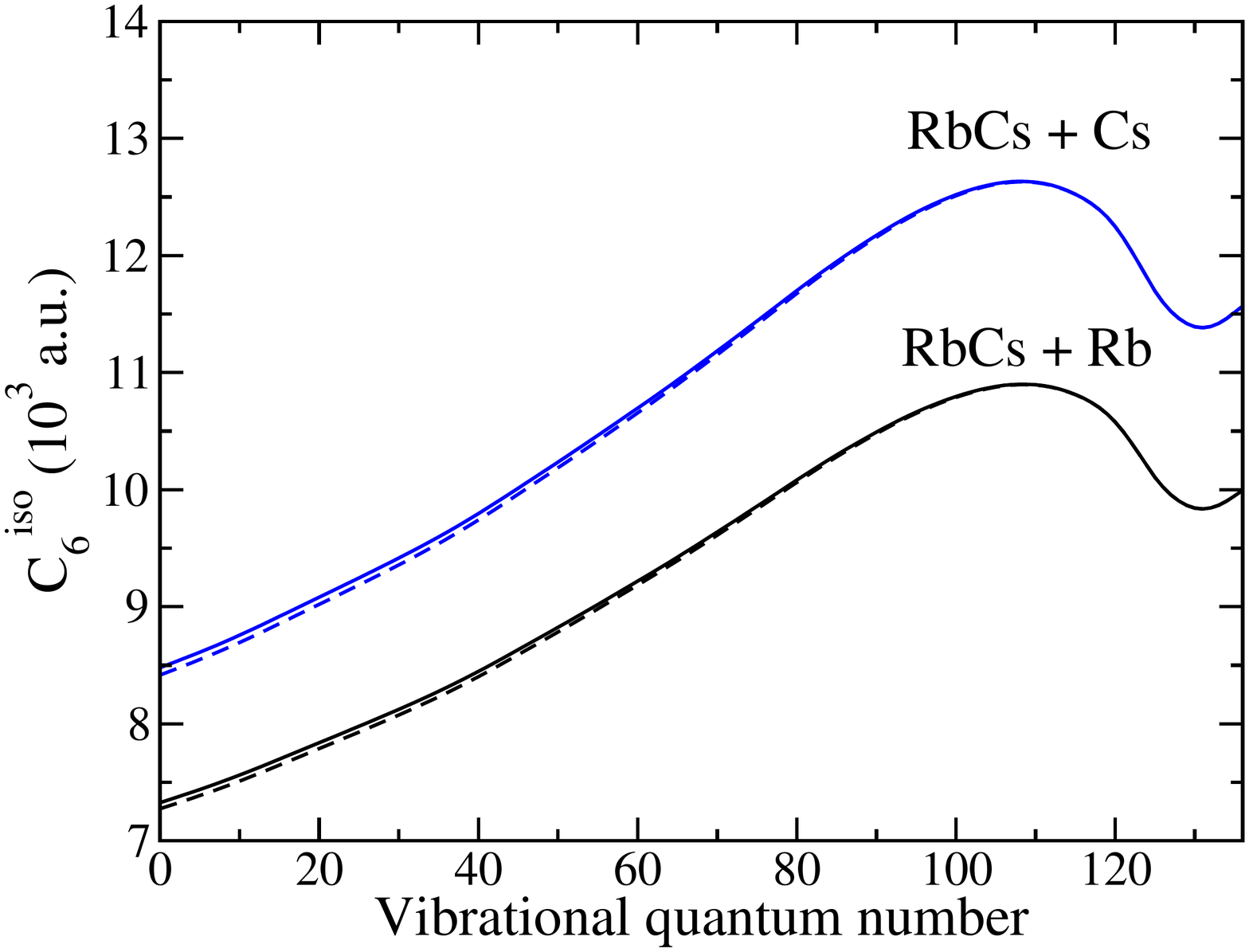}
\includegraphics[scale=0.3,viewport=0 20 750 530,clip]{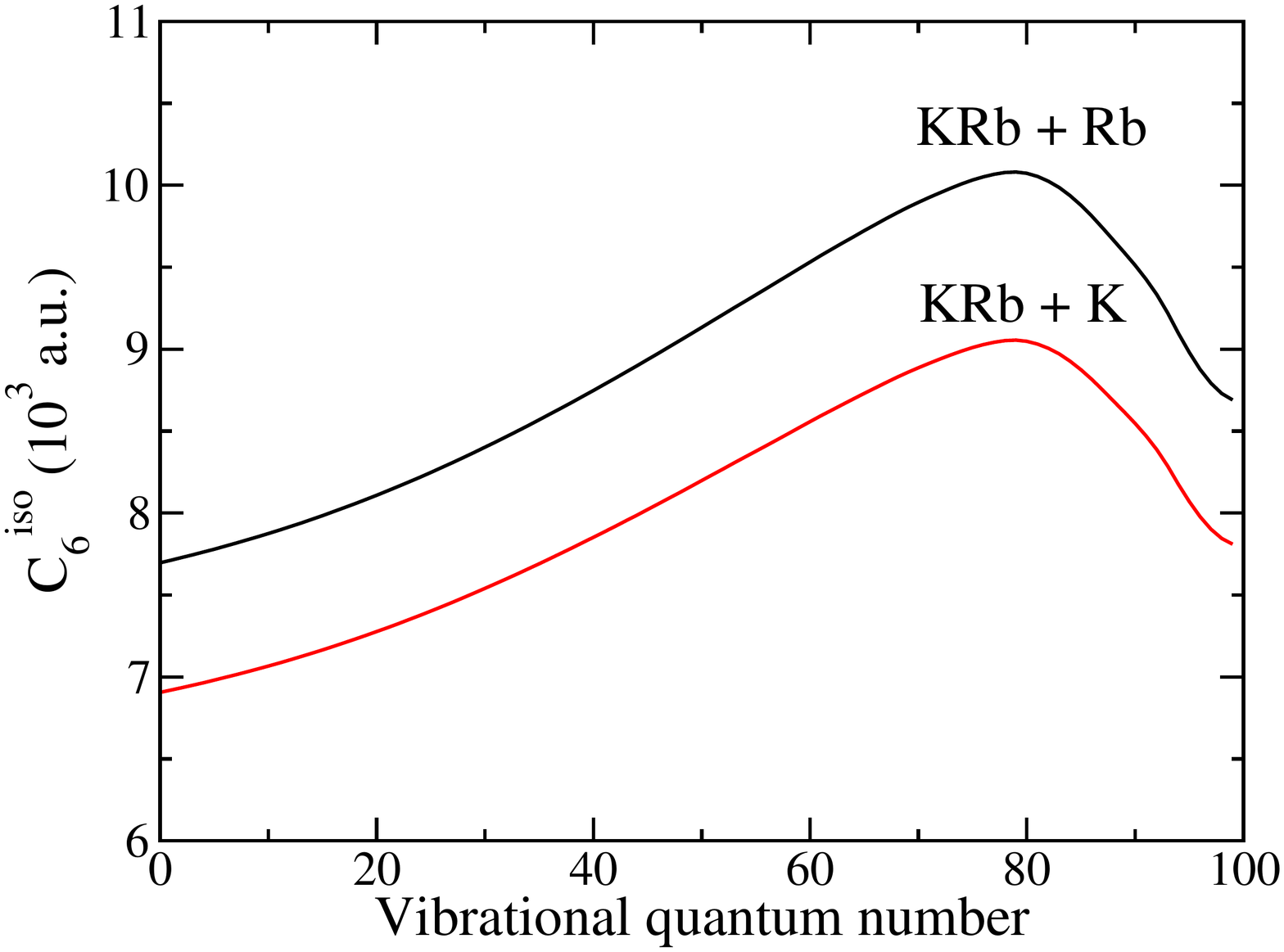}
\caption{Isotropic van der Waals coefficients in atomic units as a function of vibrational quantum number for the collision between a ground  X$^1\Sigma^+$ state
RbCs  molecule and either a Rb or Cs atom (left panel) and KRb molecule and either Rb or K atom (right panel).
Rotational J=0 (1) levels are shown as solid (dashed) lines, respectively. For KRb the J=0 and 1 $C_6^{\rm iso}$
coefficients are indistinguishable on the scale of a figure.}
\label{C6MolAtom}
\end{figure}

\begin{figure}
\includegraphics[scale=0.27]{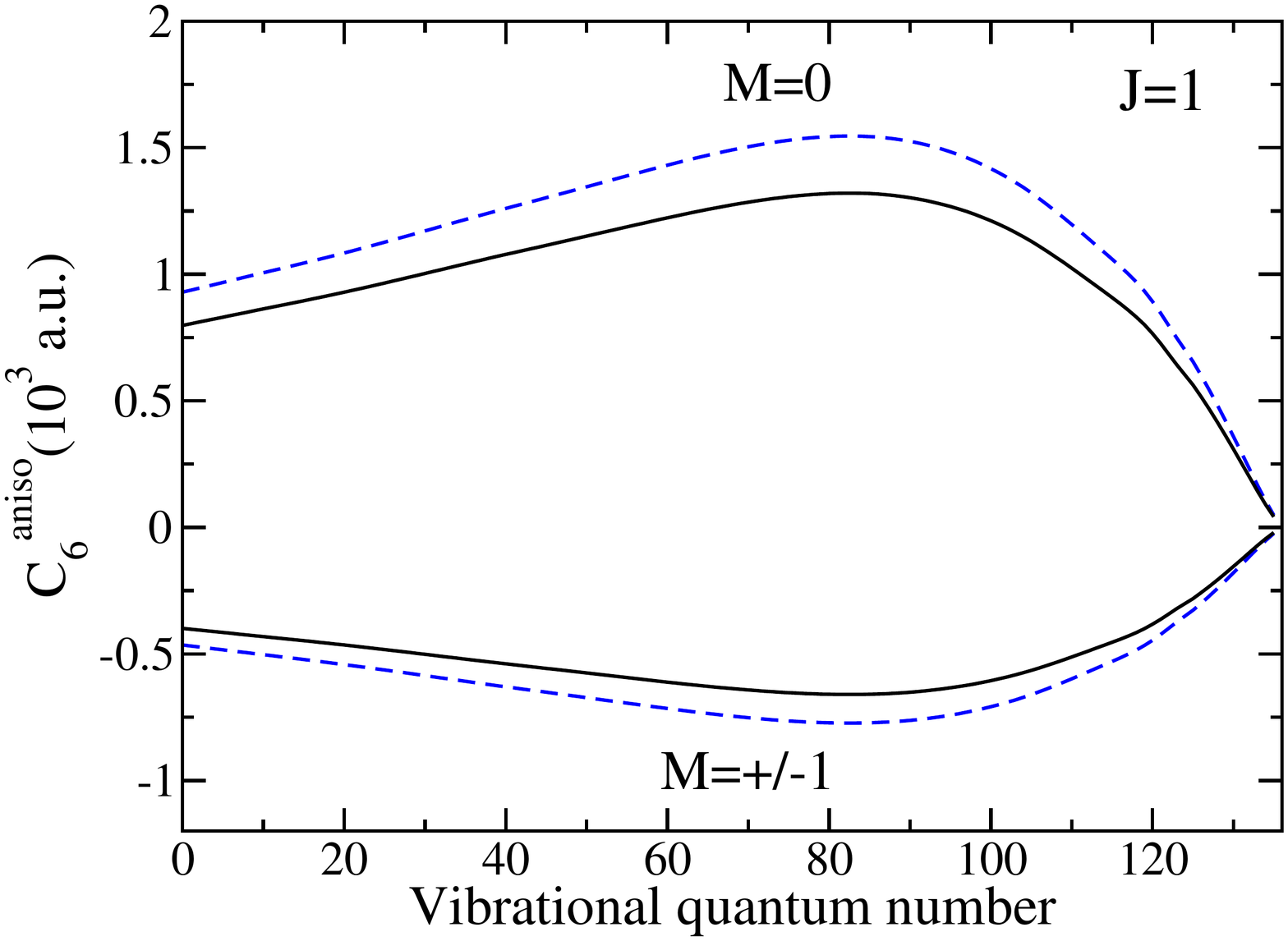}
\includegraphics[scale=0.27]{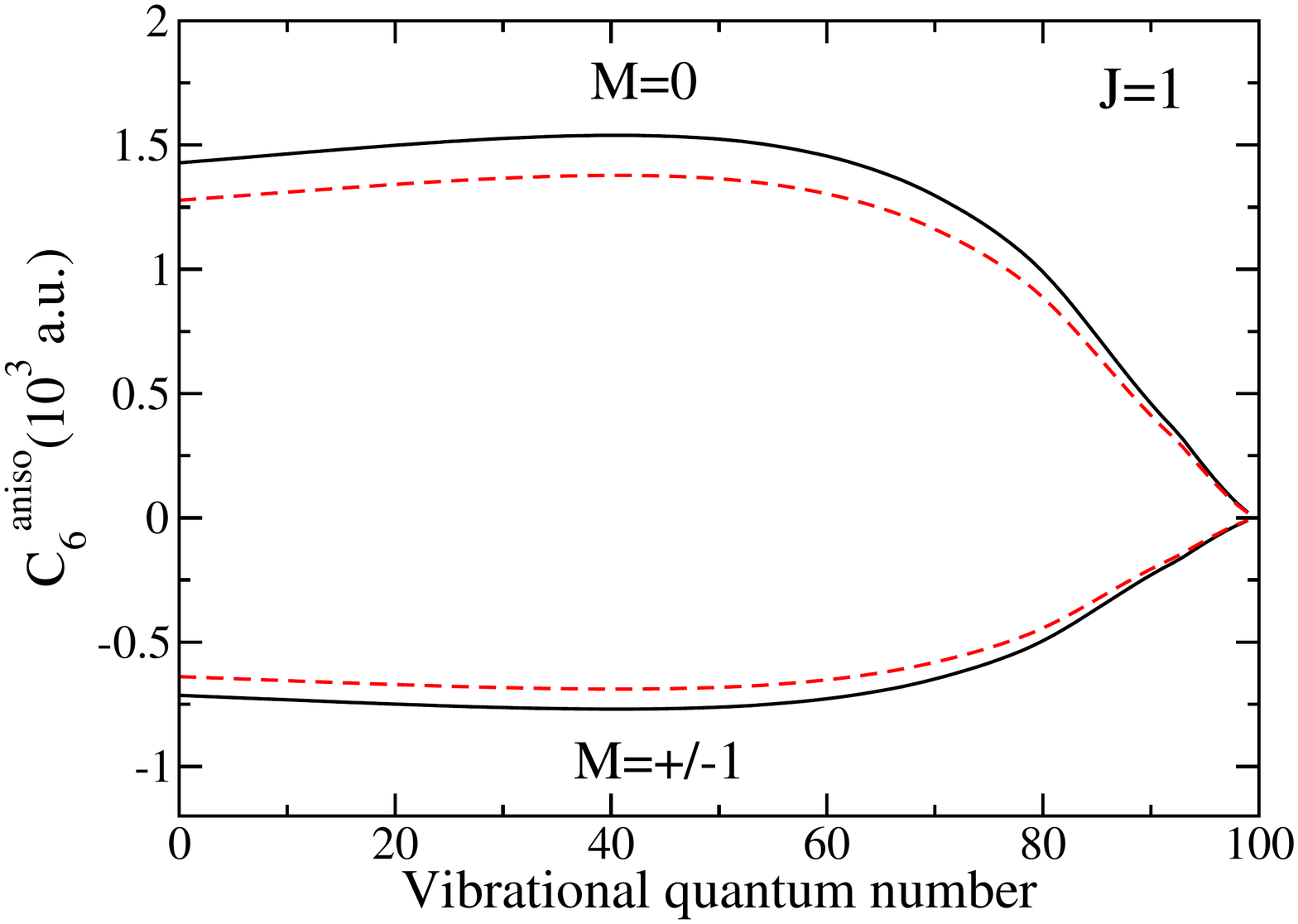}
\caption{Anisotropic van der Waals coefficients $C_{6,20}^{\rm aniso}$
in atomic units as a function of vibrational quantum number for $J=1$
rotational $M=0$ and $M=\pm 1$  sublevels of the X$^1\Sigma^+$ ground
state of the RbCs+Rb (solid lines) and RbCs+Cs (dashed lines) interacting
systems (left panel) and KRb+Rb (solid lines) and KRb+K (dashed lines)
interacting systems (right panel).}
\label{aniso_AtMol}
\end{figure}

Figure \ref{C6MolAtom} shows the isotropic van der Waals coefficients
for  RbCs and KRb molecules with constituent atoms of Cs, Rb and Rb, K, respectively.
It provides evidence  that the van der Waals coefficients for both $J$=0 and 1  are
nearly the same. This is due to the fact that the contribution from transitions
within the ground state are negligible. The lack of these contributions effects the
behavior of an anisotropic $C^{\rm AtMol,aniso}_6$ coefficient as well. As it is shown
in Fig.~\ref{aniso_AtMol}, the values of the $C^{\rm AtMol,aniso}_6$ for the same projections  
$M$ have the same sign. Table~\ref{summary} lists the isotropic and anisotropic values for 
the $v=0$ vibrational level of the X$^1\Sigma^+$ state of KRb and RbCs with atoms.

Similar to the molecule-molecule dispersion potential, the van der Waals
coefficients for one molecule and  one atom are additive
for a weakly-bound molecule. In other words, the $C_6$ coefficient
for the highly excited vibrational levels can be expressed as a sum  of
contributions from individual di-atoms. For example,
$C^{\rm RbCs+Rb,iso}_6\to C_6({\rm RbCs}) + C_6({\rm Rb}_2)  = 10035$
a.u.  and $C^{\rm RbCs+Cs,iso}_6\to C_6({\rm RbCs}) + C_6({\rm Cs}_2)
= 12514$ a.u \cite{Derevianko_C6} for weakly bound RbCs molecules.
The agreement between the asymptotic values and our numerical
calculations is satisfactory.  The same
estimate can be done for KRb molecules interacting at long distances
with K or Rb atoms.  This lead to   $C^{\rm KRb+Rb,iso}_6\to C_6({\rm KRb})
+ C_6({\rm Rb}_2)  = 8965$ a.u.  and $C^{\rm KRb+K,iso}_6\to C_6({\rm
KRb}) + C_6({\rm K}_2)  = 8171$ a.u., which agree within a few $\%$ 
with the numerical values of Fig.~\ref{C6MolAtom}. 

\begin{table}[b]
 \caption{Van der Waals coefficients in atomic units
for the interaction between two molecules in the $v=0$, $J=0$ and $1$
rovibrational levels of the X $^1\Sigma^+$ potential as well as between
such a molecule with an atom. The uncertainty in the coefficients is 5\%.} \label{summary}
\vspace*{0.2cm}
\begin{tabular}{|l|c|c|c|c|c|c|c|} \hline
\multicolumn{1}{|c|}{System} &\multicolumn{2}{c|}{$C_6^{\rm iso}$}&
        \multicolumn{2}{c|} {$C_{6,20}^{\rm aniso}$}&\multicolumn{2}{c|} {$C_{6,22}^{\rm aniso}$}\\
                  &  $J=0$  & $J=1$  &   $(J,M)=(1,0)$ & $(J,M)=(1,\pm 1)$&$(J,M)=(1,0)$ & $(J,M)=(1,\pm 1)$\\ 
\hline
${\rm KRb+KRb}$   &  16133  & 13749  &        2569     &   -1285 & -43  & -11 \\ 
${\rm RbCs+RbCs}$ & 142129  & 16865  &        1443     &   -2886 & -630 & -157 \\
${\rm KRb+Rb}$    &   7696  &  7686  &        1428     &    -714 &      &      \\
${\rm KRb+K}$     &   6905  &  6896  &        1278     &    -639 &      &      \\  
${\rm RbCs+Rb}$   &   7326  &  7274  &         798     &    -399 &      &      \\
${\rm RbCs+Cs}$   &   8479  &  8416  &         929     &    -465 &      &      \\ \hline
\end{tabular}
\end{table}

\section{Reactive collisions of ultracold polar molecules} \label{scatter}

The lifetime of ultracold molecules in an optical trap is determined by
the inelastic collisional loss rates between these molecules.  A collision
can change the internal rovibrational state of the molecules as well
induce reactions where the bonds between the atoms are rearranged. For
example, for an ultracold KRb gas the reaction KRb+KRb $\to$ K$_2$+Rb$_2$
is allowed~\cite{Science2010}.  Once the molecules are prepared in their
absolute ground state only the reactive process is present.

This section describes our results using a quantum reflection model of the rate
coefficient for rotationless $J=0$ $^{40}$K$^{87}$Rb and $^{85}$Rb$^{133}$Cs ground
state molecules.  It is based on a modification of the approach developed in
Refs.~\cite{Mies,Julienne}.  We solve a single-channel radial Schr\"odinger equation
for the inter-molecular separation with  an isotropic van der Waals potential for
$R>R_{\rm sr}$, where $R_{\rm sr}$ is a short-range separation to be defined below.
Two short-range parameters  describe the collisional wavefunction when the molecules
are ``close'' together at $R=R_{\rm sr}$.  As we will show this boundary condition
models the physics or reactivity for $R<R_{\rm sr}$ and  ``partially'' inelastic
events can occur.  The short-range parameters can depend on the rovibrational state
of the molecules.  The isotropic van der Waals coefficients are taken from
Table~\ref{summary}.  Recently, a similar model with a different treatment of the
short-range interactions was used in Ref.~\cite{MQDT2009} for the collisions between
KRb molecules.

We  use the van der Waals potential to determine the validity of the model and
estimate characteristic length scales to distinguish between the dispersion and
short-range interaction regions.  The range of the isotropic van der Waals potential
is defined as $R<R_6 \equiv \sqrt[4]{2\mu C^{\rm iso}_6/\hbar^2}$, where $\mu$ is the
reduced molecular mass \cite{Gao2000}.  For  $v=0$, $J=0$ KRb and RbCs molecules this
separation is $\approx$ 300 $a_0$ and 500 $a_0$, respectively.  The short-range
$R_{\rm sr}$ is defined by  $C^{\rm iso}_6/R_{\rm sr}^6 \approx 2 B_v$, where $B_v$
is the rotational constant of the ground state molecule.  For shorter separations
the energy of the dispersion potential is much larger than the rotational energy and
collisional interactions can mix rotational states.  Such mixing falls outside the
scope of our single channel description.  We find  values for $R_{\rm sr}$ between 50
$a_0$  and 80 $a_0$ for both molecules.  Hence $R_{\rm sr}\ll R_6$. In fact,  $C^{\rm
iso}_6/R_{\rm sr}^6$ is  much larger than  collision energies of interest as well as
the centrifugal potential $\hbar^2l(l+1)/(2\mu R_{\rm sr}^2)$ {\it between}
molecules.  The quantum number $l$ is the partial wave between the molecules or one
molecule and one atom.

Our model involves scattering of rotationless molecules in the potential $-C^{\rm iso}_6/R^6+\hbar^2l(l+1)/(2\mu R^2)$ for $R>R_{\rm sr}$.  The contribution for partial
wave $l$ and projection $m$ to the inelastic rate coefficient is given by 
\begin{equation}
    K_{lm}^{\rm loss}(E)= v\frac{\pi}{k^2} \sum_{\alpha\neq i} |S_{i\alpha}(E,lm)|^2  =  v\frac{\pi}{k^2} \left( 1- |S_{ii}(E,lm)|^2 \right)\,,
    \label{lossrate}
\end{equation} 
where $E=\hbar^2k^2/(2\mu)$ is the collision energy and $v$ is the relative
velocity.  The scattering S matrix elements $S_{i\alpha}(E,lm)$ describe the
transmission and reflection amplitude from the initial state $i$ of colliding
molecules and atoms to a  final state $\alpha$ with either molecules in different
rovibrational states or with a different bond. The sum over $\alpha$ excludes the
initial state. Flux conservation or the unitarity of the $S$ matrix allows us to
rewrite the loss rate coefficient in terms of the diagonal S matrix element,
$S_{ii}$.  At ultracold temperatures, only a few partial waves $l$ contribute  as for
higher $l$ the centrifugal barrier  prevents the molecules from approaching each
other and $K^{\rm loss}_{lm}$ rapidly goes to zero.

The scattering S matrix is calculated from the radial Schr\"odinger equation
\[
 \left(   -\frac{\hbar^2}{2\mu}\frac{d^2}{dR^2}  -\frac{C^{\rm iso}_6}{R^6}+\frac{\hbar^2}{2\mu} \frac{l(l+1)}{R^2}
    \right) \psi_{lm}(R) = E\psi_{lm}(R)
\]
at collision energy $E$ with the boundary condition
\begin{equation}
\psi_{lm}(R)=A\left(e^{+i[y -\pi/4]} - \zeta_{lm}(E) e^{2i\delta_{lm}(E)} e^{-i[y -\pi/4]} \right)
           \end{equation}
at $R=R_{\rm sr}\ll R_6$,  $y=(R/R_6)^{-2}/2$, and $A$ is a normalization constant
\cite{Gao2000}. The exponents $e^{\pm i[y -\pi/4]} $ can be recognized as  WKB
solutions of a van der Waals potential at zero collision energy.  The short range
amplitude $\zeta_{lm}(E) e^{2i\delta_{lm}(E)}$ and thus the real parameters $\zeta_{lm}(E)$
and $\delta_{lm}(E)$ are not known a priori.  They can only be determined from the
chemical bonding between the three or four atoms.  However, once they are known the
boundary condition uniquely specifies the solution of the Schr\"odinger equation.
Flux conservation requires that $0\leq \zeta_{lm} \leq 1$, where $\zeta_{lm}=0$
corresponds to the case where no flux is returned from the short range and
$\zeta_{lm}=1$ correspond to the case where everything is reflected back.  The phase
$\delta_{lm}$ describes the relative phase shift of the flux that returns from
$R<R_{\rm sr}$.  In Ref.~\cite{MQDT2009} a model potential with an optical or
imaginary term is used to describe the short range physics for $R<R_{\rm sr}$.

The van der Waals potential is the largest energy scale at $R=R_{\rm sr}$ and,
consequently, we {\it initially} assume that both $\zeta_{lm}(E)$ and $\delta_{lm}(E)$ are
independent of collision energy $E$, partial wave $l$, and projection $m$. The
validity of these approximations can only be tested by comparison with experimental
data. We will do so with the data on KRb loss rate coefficients from
Ref.~\cite{Science2010}.

In the limit $R\to\infty$ the wavefunction approaches
 \begin{equation}
              \psi_l(R)   \to e^{-i(kr-l\pi/2)} - S_{ii}(E,lm)e^{i (kr-l\pi/2)}\,,
\end{equation}
where $S_{ii}(Elm)$ is the diagonal S matrix element required for Eq.~(\ref{lossrate}).
In order to determine the relationship between $\zeta_{lm} e^{2i\delta_{lm}}$ and
$S_{ii}(E,lm)$ we solve the Schr\"odinger equation analytically in the limit $R_{\rm
sr}\to 0$ using the solutions from \cite{Gao2009}.  Alternatively, the Schr\"odinger
equation can be solved numerically.

\begin{figure}
\includegraphics[scale=0.70,viewport=10 30 330 220,clip]{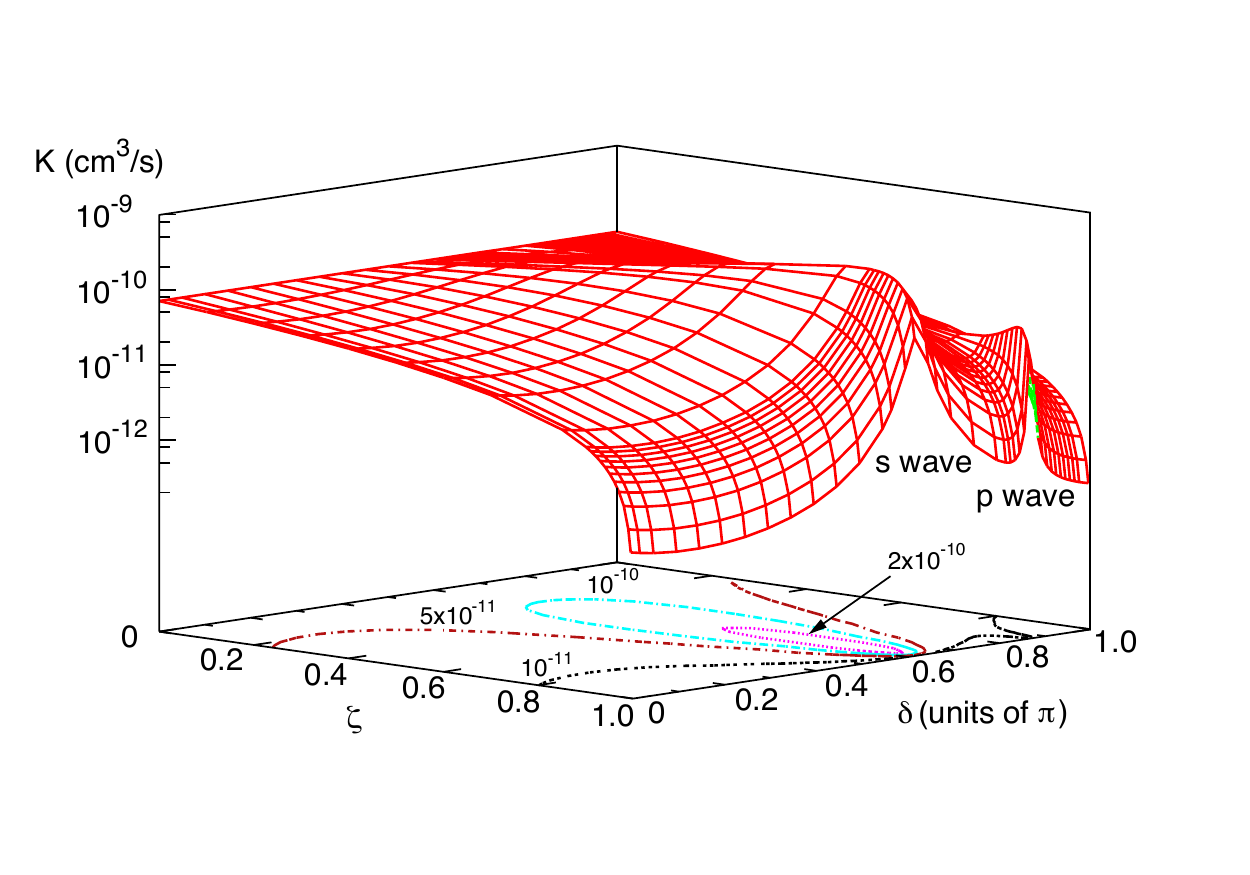}
\includegraphics[scale=0.70,viewport=10 30 330 220,clip]{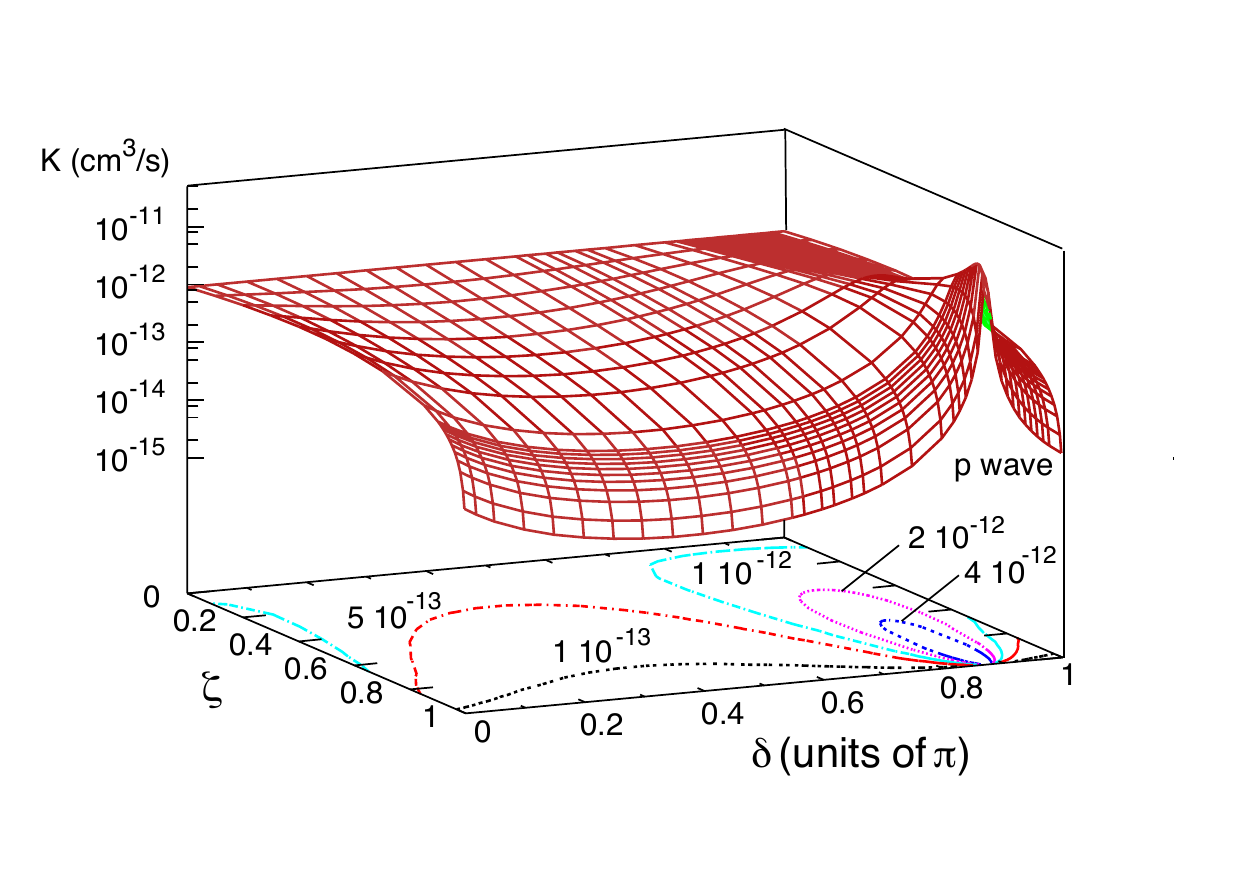}
\caption{The total inelastic loss rate coefficient for a nuclear spin unpolarized
(left panel) and polarized (right panel) sample of $v=0$, $J=0$ $^{40}$K$^{87}$Rb molecules in the X$^1\Sigma^+$
potential as a function of the short-range parameters $\zeta$ and $\delta$ at a
collision energy of $E/k_B=350$ nK. The loss rate of the unpolarized case 
contains non-negligible contributions from $s$ and $p$ wave contributions. 
The polarized case only contains odd partial waves.  }
\label{fig:lossKRb}
\end{figure}

The left panel of Fig.~\ref{fig:lossKRb} shows the total loss rate coefficient summed
over $l$ and $m$ for  collisions between fermionic $^{40}$K$^{87}$Rb molecules in the $v=0$,
$J=0$ rovibrational state of the X$^1\Sigma^+$ potential as a function of the
short-range parameters $\zeta$ and $\delta$ at a collision energy of $E/k_B=350$ nK.
For this figure we assume that $\zeta_{lm}\equiv\zeta$ and $\delta_{lm}\equiv\delta$ are
independent of $l$ and $m$.  This rate coefficient can be compared with experimental
data on an ``unpolarized'' sample of  this fermionic KRb molecule with at least two
equally-populated nuclear-hyperfine states.  The right panel shows the loss rate 
coefficient  at $E/k_B=350$ nK for a
``polarized'' sample where all molecules are in the same nuclear hyperfine state and
only odd partial wave scattering is allowed.  In fact, for this collision energy only
the $p$ wave is significant.  The $p$ wave loss rate is much smaller than that for
the $s$ wave, since the centrifugal barrier is larger than the 350 nK collision
energy and the $p$-wave loss rate is reduced.  We use $C^{\rm iso}_6=16133$ a.u.  for
both panels.

The most striking observation of Fig.~\ref{fig:lossKRb} is the appearance of
resonances in the loss rate. The origin of this behavior is in the interference
between the in- and out-going flux for $R>R_{\rm sr}$.  The maximum at
($\zeta\approx 0.8$, $\delta\approx0.6\pi$) is due to $s$-wave collisions and  that
at ($\zeta\approx 0.95$, $\delta\approx0.9\pi$) due to $p$-wave collisions.  In fact,
the loss rate is significantly larger than the rate  at $\zeta=0$ when no flux
returns from $R<R_{\rm sr}$. (For $\zeta=0$ the loss rate is independent of
$\delta$.) The loss rate can also be much smaller than at $\zeta=0$.  In the limit
$\zeta\to 1$ we have $K^{\rm loss}\to 0$. For clarity, this limiting behavior is not
shown in the surface plot.
If we learn how to change the short range parameter we can envision that a significantly
reduction of the loss rate is possible.

\begin{figure}
\includegraphics[scale=0.4,viewport=10 40 630 550,clip]{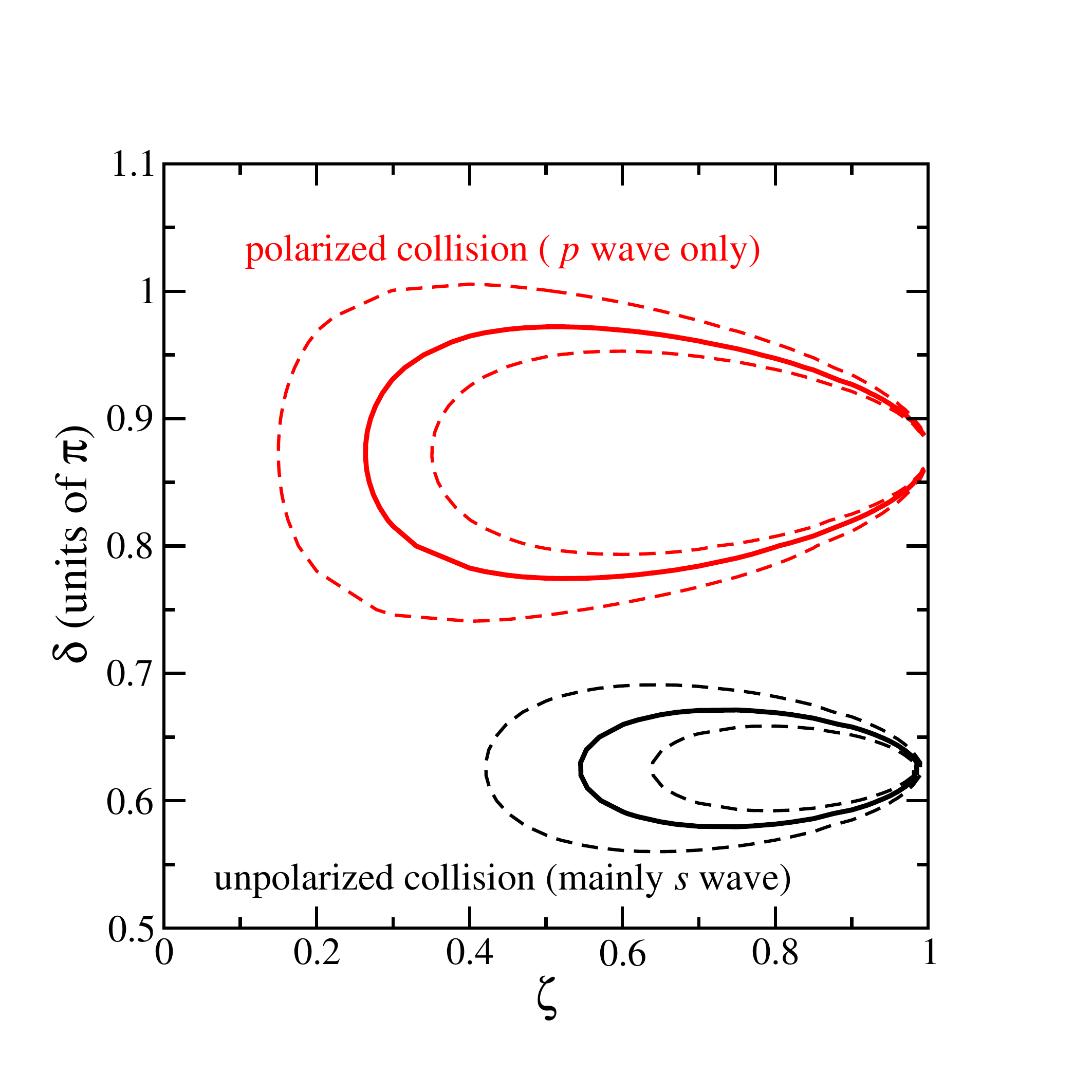}
\caption{Bounds on the allowed values of the short-range parameters $\zeta$ and
$\delta$ from the experimental loss rates obtained in Ref.~\cite{Science2010} for
either a nuclear spin unpolarized (black curves) or polarized (red curves) sample of
$v=0$, $J=0$ $^{40}$K$^{87}$Rb molecules in the X$^1\Sigma^+$ potential at a
temperature of $T=250$ nK.  The full lines correspond to pairs of $(\zeta,\delta)$
where the experimental loss rate agrees with our thermally averaged loss rate.  The
dashed lines follow from the experimental uncertainty limits.  In fact, at a
temperature of 250 nK the measured unpolarized loss rate is $\beta=1.9(4)\cdot
10^{-10}$ cm$^3$/s, while that for the polarized case is $\beta=3.3(7) \cdot
10^{-12}$ cm$^3$/s.}
\label{fig:contourKRb}
\end{figure}

Figure~\ref{fig:contourKRb} shows a comparison of thermally-averaged rate
coefficients for ground state $^{40}$K$^{87}$Rb molecules with the experimental
observations of Ref.~\cite{Science2010}.  The thermal average assumes a
Maxwell-Boltzmann distribution.  In order to obtain the comparison we not only
assumed that the $\zeta_{lm}$ and $\delta_{lm}$ are independent of partial wave but also
independent of collision energy over several $kT$, where $k$ is the Boltzmann
constant.  

The experimental rates are $\beta_{\rm u}= 1.9(4)\cdot 10^{-10}$ cm$^3$/s
for the unpolarized case and $\beta_{\rm p}=3.3(7)\cdot 10^{-12}$ cm$^3$/s for the
polarized case at a temperature of $T=250$ nK.  In the experiment $\beta$ was
extracted from the time-dependence of the molecule number density of each nuclear
spin state.  Consequently, our theoretical loss rate has been multiplied by a factor
two for a polarized sample to account for the loss of two molecules per inelastic
collision, and multiplied by one(1) for the unpolarized sample as one atom of each
nuclear hyperfine state is lost.
Ref.~\cite{Science2010} verified that the rates for the polarized and unpolarized
sample satisify their $\beta_{\rm u}\propto {\rm const.}$ and $\beta_{\rm p}\propto
kT$ Wigner threshold limits as a function of temperature upto $T\approx 1$ $\mu$K.

The figure shows regions in the plane $(\zeta,\delta)$, where the theoretical rate
coefficient agrees with the experimental rate coefficients within the uncertainties.
One region is for the $p$-wave dominated polarized case and one for the $s$-wave
dominated unpolarized case.  The two regions do no overlap. Hence, our initial
assumption that the short-range parameters are independent of partial wave $l$ is
not valid. The validity of the Wigner threshold limit over collision energies upto
a few $\mu$K suggests that the short-range parameters are nevertheless independent of
energy over this range. In other words the behavior at $R<R_{\rm sr}$, where a
multi-channel description is needed to describe the reactive rearrangement of the
atoms, must be captured by more than two parameters.  The uncertainty in the
dispersion coefficient does not modify this conclusion.

\begin{figure}
\includegraphics[scale=0.75,viewport=0 30 330 220,clip]{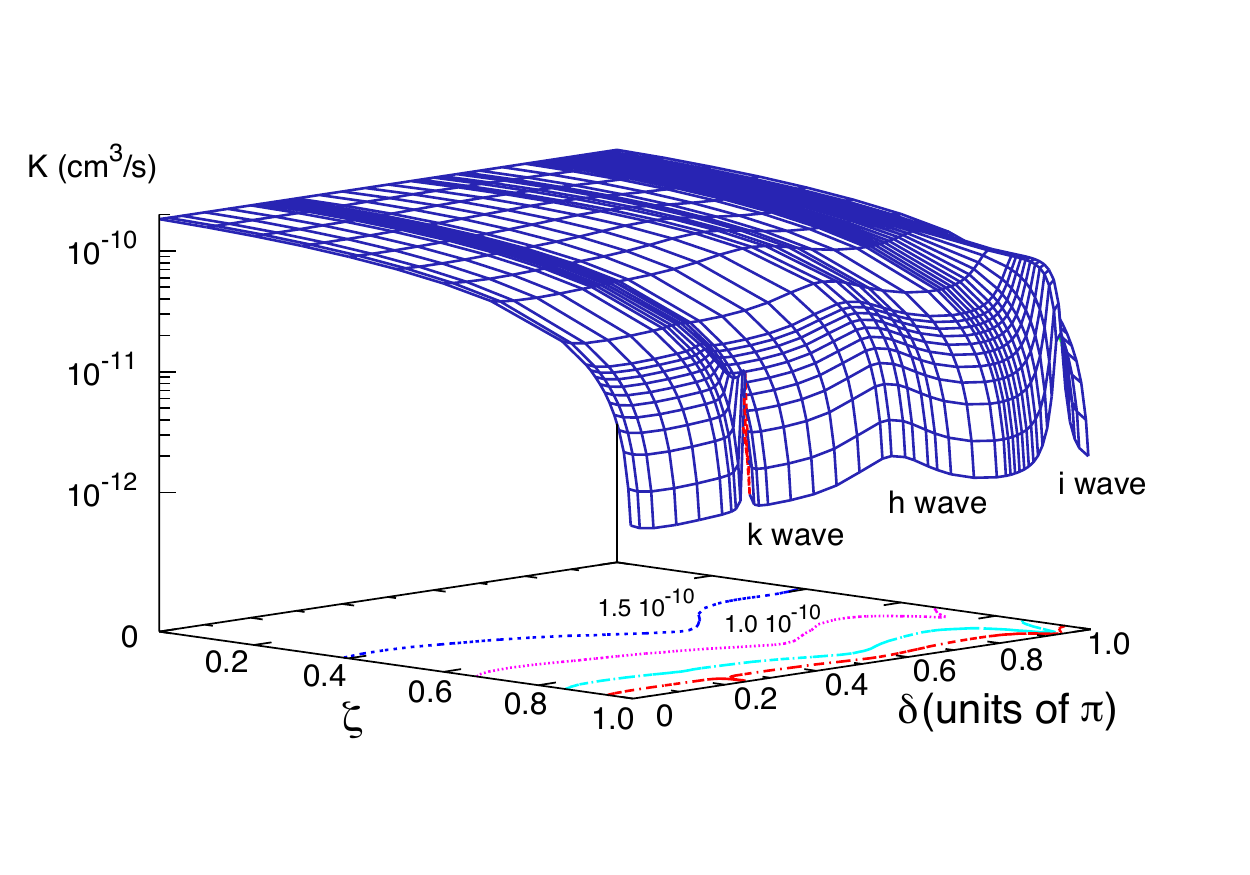}
\caption{The total inelastic loss rate coefficient for an unpolarized  sample of
$v=1$, $J=0$ $^{87}$RbCs molecules in the X$^1\Sigma^+$ potential as a function of
the short-range parameters $\zeta$ and $\delta$ at a collision energy of
$E/k_B=250$ $\mu$K.  At this collision energy many partial waves contribute.}
\label{fig:lossRbCs}
\end{figure}

Figure~\ref{fig:lossRbCs} shows the loss rate coefficient for the collision between
two bosonic $^{87}$RbCs molecules in the vibrationally excited $v=1$, $J=0$ level of the X$^1\Sigma^+$
potential at a collision energy of $E/k=250$ $\mu$K and $C_6^{\rm iso}=142540$ a.u.
The nuclear hyperfine state are equally populated and partial waves up to $l=8$ are
included.  We again assume that the short-range parameters are independent of partial
wave.  In the figure maxima or resonances for $l=5$, 6, and  7 are observed.  
As can be found from a comparison of the binding energies of the $v=0$ vibrational
levels the three RbCs, Rb$_2$,
and Cs$_2$ dimer molecules, chemical reaction at ultracold temperature 
RbCs+RbCs$ \rightarrow$ Cs$_2$+Rb$_2$ is endothermic by $\approx$ 40 cm$^{-1}$ and not 
energetically allowed. For collisions between $v=1$ RbCs molecules the reaction is exothermic.
At a 250 $\mu$K collision energy rate coefficients are as large as $2 \cdot 10^{-10}$ cm$^3$/s can be
seen in Fig.~\ref{fig:lossRbCs}. With typical densities in ultracold experiments 
around $10^{11}$ 1/cm$^3$ to  $10^{12}$ 1/cm$^3$ lifetimes as short as a 10 ms could be
observed.

\section{Conclusion} \label{last}

We have performed a theoretical study critical for the physical
realization of highly controlled ultracold molecular systems.  In
particular, we focused on hetero-nuclear KRb and RbCs molecules
used in ongoing ultracold experiments.
Our detailed calculations of C$_6$ dispersion coefficients for
interactions between molecules or molecules with atoms helped 
the experimental \cite{Science2010} and theoretical \cite{MQDT2009} 
efforts to quantitatively describe quantum-state controlled chemical
reactions.

Analyses of the isotropic and anisotropic interaction
potentials of ultracold polar molecules unveiled a significant contribution
from dipole coupling to electronically excited states. This leads to
a dramatic change in the interaction potentials as compared to previous
estimates based solely on the permanent dipole moment.

Our scattering calculations predict constructive and destructive
interferences in the molecular scattering loss rates as a function of
short-range parameters. Comparison to recent experimental measurements
\cite{Science2010} shows that the initial assumption that the short-range
parameters are independent of partial wave $l$ is not valid.

\section{Acknowledgements}

This work is supported by Air Force Office of Scientific Research MURI on Ultracold Molecules. 
The author acknowledges helpful discussions with Piotr Zuchowski and Eite Tiesinga.

\end{document}